\documentclass[12pt,a4paper]{article}

\usepackage[utf8]{inputenc}
\usepackage[nottoc]{tocbibind}

\usepackage{amsmath,amssymb}				
\usepackage{amsfonts}				
\RequirePackage{bbm}
\usepackage[thref,amsmath,thmmarks]{ntheorem}	
\usepackage{prettyref}				
\usepackage{multicol}
\usepackage{paralist,epsfig}				
\usepackage{tikz}
\usetikzlibrary{calc,intersections,arrows}

\usepackage{environ}
\makeatletter
\newsavebox{\measure@tikzpicture}
\NewEnviron{scaletikzpicturetowidth}[1]{%
  \def\tikz@width{#1}%
  \def\tikzscale{1}\begin{lrbox}{\measure@tikzpicture}%
  \BODY
  \end{lrbox}%
  \pgfmathparse{#1/\wd\measure@tikzpicture}%
  \edef\tikzscale{\pgfmathresult}%
  \BODY
}
\makeatother

\usepackage{url}
\usepackage[colorlinks=true]{hyperref}
\usepackage{tensor}
\usepackage{lmodern}
\usepackage{textcomp}
\usepackage{graphicx}

\newcommand {\bC}{{\mathbb C}}

\newcommand {\bN}{{\mathbb N}}

\newcommand {\bR}{{\mathbb R}}

\newcommand {\cF}{{\mathcal F}}
\newcommand {\cG}{{\mathcal G}}
\newcommand {\cH}{{\mathcal H}}

\newcommand {\cK}{{\mathcal K}}

\newcommand {\cO}{{\mathcal O}}

\newcommand {\cT}{{\mathcal T}}
\newcommand {\cV}{{\mathcal V}}
\newcommand {\cX}{{\mathcal X}}

\newcommand {\bsd}{\boldsymbol{d}}
\newcommand {\bsi}{\boldsymbol{i}}

\newcommand {\Sing}{\operatorname{Sing}}
\newcommand {\Wand}{\operatorname{Wand}}

\newcommand {\Trans}{\operatorname{Trans}}
\newcommand {\Set}{\operatorname{Set}}

\newcommand{\1}{\mathbbm 1}

\newcommand{\rstr}{{\upharpoonright}}

\theoremnumbering{arabic}

\theoremstyle{plain}
\RequirePackage{latexsym}
\theorembodyfont{\itshape}
\theoremheaderfont{\normalfont\bfseries}
\theoremseparator{}

\newtheorem{Theorem}{Theorem}[section]

\newtheorem{Lemma}[Theorem]{Lemma}
\newtheorem{Corollary}[Theorem]{Corollary}

\newtheorem{Example}[Theorem]{Example}
\newtheorem{Remark}[Theorem]{Remark}

\newtheorem*{TheoremN}{Theorem}

\theoremstyle{nonumberplain}
\theorembodyfont{\upshape}

\theoremheaderfont{\itshape}
\theorembodyfont{\normalfont}
\theoremsymbol{\ensuremath{_\Box}}
\RequirePackage{amssymb}
\qedsymbol{\ensuremath{_\Box}}

\theoremclass{LaTeX}

\numberwithin{equation}{section}

\usepackage{color}

\newcommand {\eh}{{\textstyle \frac{1}{2}}}

\begin{document}

\title {Improbability of Wandering Orbits Passing Through a Sequence of Poincar\'e Surfaces of Decreasing Size}
\author{Stefan Fleischer and Andreas Knauf\thanks{
Department of Mathematics,
Friedrich-Alexander-University Erlangen-N\"urnberg,
Cauerstr.\ 11, D-91058 Erlangen, 
Germany, \texttt{fleischer,knauf@math.fau.de}} 
}
\date{\today}
\maketitle
\begin{abstract}
Given a volume preserving dynamical system with non-compact phase space, one is 
sometimes interested in special subsets of its wandering set.
One example from celestial mechanics is the set of initial values leading to collision.
Another one is the set of initial values of semi-orbits, whose asymptotic velocity
does not exist as a limit.  
We introduce techniques that  can be helpful in
showing that these sets are of measure zero: 
by defining a sequence of hypersurfaces, that are eventually hit by 
each of those semi-orbits and whose total surface area decreases to zero.
\end{abstract}
\section{Introduction and Main Result}\label{sec:Intro}
Let $P^d$ be a smooth  manifold with a volume form $\Omega$ and a $C^1$ 
vector field $X:P\to TP$, so that the Lie derivative ${\cal L}_X\Omega$ vanishes.
By standard results of ordinary differential equations, the flow $\Phi$
associated to the differential equation $\dot x = X(x)$
uniquely exists on a maximal neighborhood 
$D\subseteq \bR \times P$ of $\{0\} \times P$ in extended phase space,
$\Phi\in C^1(D,P)$, and $\Phi$ preserves the volume form $\Omega$. 

The flow's domain of definition is of the form
\begin{equation}
\label{eq:DefD}
D = \left\{ (t,x) \in \bR \times  P \ \left| \  T^-(x) < t < T^+(x) \right.\right\}
\end{equation}
with the so-called {\em escape time} $T:=T^+: P \rightarrow (0,+\infty]$, which is a lower semi-continuous function. Similarly, $T^-: P \rightarrow [-\infty,0)$ is upper semi-continuous.
By $\cO(x) := \Phi\left( (T^-(x),T^+(x) ), x \right)$ we denote the orbit passing through 
$x\in P$, and by $\cO^+(x) := \Phi\left([0,T^+(x) ), x \right)$ the semi-orbit.

We consider the {\em wandering set} of $\Phi$
\[\Wand\equiv \Wand_\Phi \subseteq P,\] 
consisting of those $x\in P$ which have a neighborhood   $U_x$ 
so that for a suitable time $t_-\in(0,T(x))$
\begin{equation*}
U_x\cap \Phi\big(\left((t_-,T(x))\times U_x\right)\cap D\big)=\emptyset.
\end{equation*}
\begin{Remark}[Wandering set] \quad\label{rem:wand}\\[-6mm]
\begin{enumerate}[1.]
\item 
In view of applications (see Example \ref{ex:coll} below) 
we allowed for finite escape times $T^\pm(x)$.
So $\Phi$ does not in general define an $\bR$-action on $P$.
In this sense our notion of
'wandering set' is a generalization of the usual definition.
\item 
As $T$ is lower semi-continuous, the set of {\em singular points}
\begin{equation}
\label{def:sing}
\Sing := \{ x \in P \ | \ T(x) < \infty\},
\end{equation}
is a Borel set.
It is wandering for the same reason, see Lemma \ref{lem:sing:wand} below.
\item 
Trivially, equilibrium points are nonwandering, so that
\begin{equation*}
 \Wand \subseteq \{ x \in P \ | \ X(x) \neq 0 \}.
\end{equation*}
The latter is an open submanifold of $P$ because of $X\in C^1(P,TP)$. 
Thus, we  assume the vector field $X$ to be non-vanishing on $P$ from the 
outset, without loss of generality. \hfill $\Diamond$
\end{enumerate}
\end{Remark}
\begin{Lemma}  \label{lem:sing:wand}
$\Sing\subseteq \Wand.$
\end{Lemma}
\textbf{Proof:}
Let $x\in\Sing$, so that $T(x)\in(0,+\infty)$. For any small $\epsilon_1>0$
there exists a flow-box chart (see also Lemma \ref{lem:FlowBox})
$\varphi:U_1 \rightarrow (-\epsilon_1,\epsilon_1) \times W_1 \subseteq \bR \times \bR^{2n-1}$
with $\varphi(x)=(0,0)$ that is reentered only a finite number of times by $\cO^+(x)$. 
Inside $U_1$, for any small $\epsilon_2>0$ there is a compact
neighborhood $U_2$ of $x$ with 
$\varphi(U_2)= [-\epsilon_2,\epsilon_2] \times W_2$ that is not reentered 
at all by $\cO^+(x)$. By a compactness argument, for any $\epsilon_3\in (0,\epsilon_2/2)$ there is a compact
neighborhood $U_3\subseteq U_2$ of $x$ with $\varphi(U_3)= [-\epsilon_3,\epsilon_3] \times W_3$ and
$\Phi([2\epsilon_2,T(x)-\epsilon_3]\times U_3)\cap U_2=\emptyset$.
Then also $\Phi((\{t\}\times U_3)\cap D)\cap U_3=\emptyset$ for all $t\in [2\epsilon_2,T(x))$, so that
$x$ is wandering.
\hfill $\Box$\\[2mm]
Now let 
\[\overline{\iota}_m: \overline{\cH}_m \rightarrow P\qquad (m\in \bN)\] 
be a sequence of (pairwise disjoint) codimension one closed
$\partial$-submanifolds\,\footnote{We call (sub)-manifolds with boundary \ 
{\em $\partial$-(sub)-manifolds}.} of $P$, which we will call {\em Poincar\'e surfaces} 
for reasons explained below.\\[2mm] 
{\bf Assumptions:}
\begin{enumerate}[1.]
\item 
The vector field $X$ is transversal to their relative interior ${\iota}_m: \cH_m\to P$.
Thus ($\bsi$ being the inner product) the $(d-1)$-form 
\[ {\cal V} := \bsi_{X} \Omega\]
on $P$ induces the volume forms ${\cal V}_m:={\iota}_m^* {\cal V}$ on $\cH_m$.
\item 
We assume that the $\cH_m$ are of finite volume: 
$\int_{\cH_m} {\cal V}_m < \infty \quad (m\in \bN)$, and
the volumes go to zero: 
\begin{equation}
\label{eq:AreaPoinSurf}
\lim_{m\to\infty}\int_{\cH_m} {\cal V}_m =0.
\end{equation}
\end{enumerate}
Then we denote the set of {\em transition points}, whose forward 
orbits eventually hit all of these surfaces by
\begin{equation}
\Trans\equiv \Trans_\Phi:=
\{x\in P\mid\exists\, m_0\in\bN\;\forall \,m\ge m_0: \cO^+(x)\cap \overline{\cH}_m\neq\emptyset \}.
\label{trans:in:sing}
\end{equation}
Our main result is the following.
\begin{TheoremN}\hspace*{-1mm}{\hypertarget{thmA}{\bf A}} 
\label{thm:Trans}\
From the assumptions  it follows that $\ \Omega(\Trans\cap\Wand) = 0$.
\end{TheoremN}
\begin{Remark}[Theorem \hyperlink{thmA}{A}] \quad\\[-6mm]
\begin{enumerate}[1.]
\item 
A volume form on a manifold 
defines a positive linear functional on the vector space of 
continuous, compactly supported functions.
Thus by the Riesz representation theorem, it induces 
a measure on its Borel sets,
and like in Theorem~{\hyperlink{thmA}{A}} we do not in 
general make a distinction between these
notions and use the same symbol. 
So instead of $\int_{\cH_m} {\cal V}$ we write ${\cal V}(\cH_m)$, etc.
\item 
The volume form also defines an orientation of the given (sub-)manifold.
\item 
In general $\Omega(\Trans)>0$ (e.g.\ for every ergodic flow on
a closed manifold $P$).
\item 
Clearly in general $\,\Omega(\Wand)$ is positive, too.
\hfill $\Diamond$
\end{enumerate}
\end{Remark}
\begin{Remark}[Cases of symplectic and K\"ahler structures] \label{rem:additional}\quad\\
To apply our result, we make use of additional structures of our volume preserving
dynamical system:
\begin{enumerate}[1.]
\item 
An important subclass is the one of Hamiltonian systems
$(P^{2n},\omega,H)$, with 
\[\Omega:=\textstyle{\frac{(-1)^{\lfloor n/2 \rfloor}}{n!}}\omega^{\wedge n},\] 
$H\in C^2(P,\bR)$ and $X\equiv X_H$ uniquely given by the equation 
$\bsi_{X_H} \omega = \bsd H$ (with the exterior derivative $\bsd$).

In this case we obtain useful expressions for the volume forms ${\cal V}_m$ in terms of
the symplectic form $\omega$.
\item 
As in Hamiltonian systems $H$ is a constant of the motion, we may restrict the maximal flow to the energy
surfaces $\Sigma_E:=H^{-1}(E)$ $(E\in \bR)$. By our assumption, $dH$ is non-vanishing. 
So the $\imath_E: \Sigma_E\to P$ are
submanifolds. It is well-known, that there exists a $(2n-1)$-form $\sigma$ on $P$
with $\bsd H\wedge \sigma=\Omega$. Although $\sigma$ is not unique, its pullbacks
$\sigma_E:=\imath_E^*\sigma$ to $\Sigma_E$ are uniquely defined volume forms  
\cite[Theorem 3.4.12]{Abraham_Marsden_1978}. 
These are invariant under the restricted flow.
We denote by $\Wand_E$ the set of wandering points and by $\Trans_E$ the
set of transition points on $\Sigma_E$ w.r.t.\ a sequence of hypersurfaces 
$\cH_m^E \in \Sigma_E$. Applying Theorem \hyperlink{thmA}{A} to
$P=\Sigma_E$ hence yields $\sigma_E(\Wand_E \cap \Trans_E) = 0$, if $\lim_{m\to\infty} \int_{\cH_m^E}
\bsi_{X_H} \sigma_E = 0$.
\item  
Finally, the symplectic manifold $(P,\omega)$ may be K\"ahler, 
that is, equipped with a Riemannian metric $g$ and complex structure $J$, so that 
$\omega(X,Y) := g(JX,Y)$.
This then allows to use simple estimates for the volume forms in terms of Riemannian
volumes.
\hfill $\Diamond$
\end{enumerate}
\end{Remark}

Considering motion on an energy surface $\Sigma_E\subseteq P$
of a K\"ahler manifold $P^{2n}$, we can modify Assumption 2.:
\begin{enumerate}[2'.]
\item 
We assume that the hypersurfaces $\cH_m\subseteq \Sigma_E$ 
are of finite  volume w.r.t.\ the  Riemannian volume form $d\cH_m$:
$\int_{\cH_m} d\cH_m < \infty \quad (m\in \bN)$, and
the volumes go to zero: 
\begin{equation*}
\lim_{m\to\infty}\int_{\cH_m} d\cH_m=0.
\end{equation*}

\end{enumerate}
\begin{TheoremN}\hspace*{-1mm}{\hypertarget{thmB}{\bf B}} \ 
On any energy surface $\Sigma_E\subseteq P$ in a K\"ahler  manifold 
it follows from Assumptions 1.\ and 2'.\ 
for the Hamiltonian flow that $\ \sigma_E(\Trans_E\cap\Wand_E) = 0$. 
\end{TheoremN}
\begin{Remark} [Comparison with Theorem \hyperlink{thmA}{A}]\quad\\ 
In the case of energy surfaces $\Sigma_E\subseteq P$ of a symplectic 
manifold $(P^{2n},\omega)$, the natural invariant volume form on 
a hypersurface $\imath:\cH\to \Sigma_E\,$ is 
$\ \imath^*\omega^{\wedge n-1}$.\,\footnote{or a constant multiple of it.}
This would be the ${\cal V}$ entering in \eqref{eq:AreaPoinSurf}.
Although Assumption 2.', using 
the Riemannian volume form $d\cH_m$, is stronger than Assumption 2.,
it may be easier to check. \hfill $\Diamond$
\end{Remark}

We give a simple example.
\begin{Example}[Collision of two particles is improbable for $n\geq2$] 
\quad\label{ex:coll}\\ 
By reduction we can model the motion by the Hamiltonian function
\[H:T^*(\bR^n\!\setminus\!\{0\})\to \bR\quad,\quad H(q,p)=\eh \|p\|^2+ V(q)\]
with a potential $V\in C^2(\bR^n\!\setminus\!\{0\},\bR)$. We assume that 
for some $\alpha\in (0,2)$, $c>0$
\[ | V(q)|\leq \frac{c}{\|q\|^\alpha}
\qquad(\|q\|\leq 1) \]
and, say $\,\lim_{\|q\|\to \infty}V(q)=0$.
Then for all $x_0\in \Sing$ we have $\lim_{t\to T(x_0)} q(t,x_0)=0$.
So all singular points are collision points.

To show that $\sigma_E(\Sing_E)=0$ for all energy surfaces $\Sigma_E$, we first
tacitly delete the rest points $(q,p)\in \Sigma_E$ 
(with $V(q)=E$, $\nabla V(q)=0$ and $p=0$),
see Remark \ref{rem:wand}.3.
Then we define the $\partial$-hypersurfaces in $\Sigma_E$ by
\begin{equation}
\label{def:H:example}
\overline{\cH}_m:= \{(q,p)\in \Sigma_E\mid \|q\|=1/m, \langle p,q\rangle \leq 0\}
\qquad(m\in \bN).
\end{equation}
It is clear that all singular points are in $\Trans_E$ and
$\sigma_E(\overline{\cH}_m)<\infty$. Also, the ${\cH}_m$ are transversal to the flow.
For $n=1$ dimension $\overline{\cH}_m$ consists of two points and thus has
$m$-independent volume, clearly violating Condition \eqref{eq:AreaPoinSurf}. 
To show \eqref{eq:AreaPoinSurf} for $n\geq 2$, we note that 
on $\overline{\cH}_m$ we have 
\[ \|q\|= \textstyle{\frac{1}{m}} \quad\mbox{and} \quad
\|p\|=\sqrt{2(E-V(q))}\leq \sqrt{2(E+c\, m^{\alpha})}\leq \sqrt{c}\,m^{\alpha/2}\]
for all sufficiently large $m\in \bN$.
The symplectic manifold $(T^*(\bR^n\!\setminus\!\{0\}),\omega_0)$ 
is K\"ahler. 
Using Theorem \hyperlink{thmB}{B}, for $\cF_m:=\{q\in \bR^n\mid \|q\|=1/m\}$
\begin{align}
\label{int:formula}
\int_{{\cH}_m}& d\cH_m =\ \eh {\rm vol}(S^{n-1})\\
&\int_{\cF_m}  (2(E-V(q))^{(n-2)/2}
\sqrt{2(E-V(q))+\textstyle{\left\| Q(q)\nabla V(q) \right\|^2}} d\cF_m(q)\,,
\nonumber
\end{align}
with the Riemannian volume element $d\cF_m$. 
Here $Q(q)$ is the projection along the direction $q/\|q\|$.
The expression is derived as follows:
\begin{enumerate}[$\bullet$] 
\item
The factor ${\rm vol}(S^{n-1})$ in \eqref{int:formula} 
is the Riemannian volume of a unit sphere in momentum space.
\item 
The factor $1/2$ is due to the condition  $\langle p,q\rangle\leq 0$ in the definition \eqref{def:H:example}
of the surface  $\overline{\cH}_m$.
\item 
At the point $q$  on the sphere $\cF_m$ of radius $1/m$, the corresponding half-sphere
im momentum space has radius $\sqrt{2(E-V(q))}$. For  $\nabla V(q)$ parallel to $q$ 
this would lead to a Riemannian
volume element $\left(2(E-V(q))\right)^{(n-1)/2}d\cF_m(q)$.
\item 
By Pythagoras, this is multiplied by the ratio
\[  \left(1+ \frac{{\left\|Q(q)\nabla V(q)\right\|^2}}{2(E-V(q))}\right)^{1/2}\]
between hypotenuse and adjacent side for the direction of the gradient.
\end{enumerate}

If we additionally assume that the force is
radial, that is $Q\nabla V(q)=0$, 
then it follows that the singular set has measure zero.

On the other hand, it is well known that collision occurs with positive measure if
$V(q)=-\|q\|^{-\alpha}$ and $\alpha\ge2$. So in the example the criterion is optimal.
\end{Example}

If $(P,\omega) = (T^*M,\omega_0)$, with the canonical symplectic form $\omega_0$ on the cotangent bundle
of a Riemannian manifold $(M,h)$
and if the Hamiltonian function is of the form 'kinetic+potential', then there is a useful integration formula,
see Theorem~\hyperlink{thmC}{C}. This can be applied to our example.

\begin{Example}
If we instead use Theorem \hyperlink{thmA}{A}, employing Theorem  \hyperlink{thmC}{C}, we obtain,
without assuming that  $Q\nabla V(q)=0$, 
\[\int_{{\cH}_m} \frac{\omega^{\wedge n-1}}{(n-1)!} =v_{n-1}\int_{\cF_m} (2(E-V(q)))^{(n-1)/2}\,d\cF_m(q).\]
The integral is bounded above by $c_3m^{(\alpha-2)(n-1)/2}$ and thus goes to zero as 
$m\to \infty$. So the singular set has measure zero.
\hfill $\Diamond$
\end{Example}
\begin{Remark}[Transversality]\quad\\[-6mm] 
\begin{enumerate}[1.]
\item 
It was important in the example to use {\em closed} $\partial$-submanifolds
$\overline{\cH}_m$ to ensure that every collision orbit belongs to $\Trans$.
On the other hand, the proof of Theorem~\hyperlink{thmA}{A} will be based
on the transversality of the flow w.r.t.\ the Poincar\'e surfaces ${\cH}_m$.
So it will be necessary to show in general that the orbits meeting $\partial {\cH}_m$
do not contribute to $\,\Omega(\Trans\cap\Wand)$. This will be done in 
Section~\ref{sec:transversality}.
\item 
In Example \ref{ex:coll} we defined for every energy surface $\Sigma_E\subseteq P$ 
of the Hamiltonian flow  $\Phi: D\to P$
hypersurfaces $\cH^E_m\subseteq \Sigma_E$ that are transversal to the flow.

Instead, we could define from the outset hypersurfaces $\cH_m\subseteq P$.
This does not imply the aimed-for result for all energy surfaces intersected by the $\cH_m$.
However, by Sard's Theorem one could conclude in this approach that
$\lambda^1$-{\em almost all} of these $\Sigma_E$ are met transversally, 
see Remark \ref{rem:trans:Sard}.
\hfill $\Diamond$
\end{enumerate}
\end{Remark}

The outline of the article is as follows: 
Section \ref{sec:Disc} abstractly considers a discretized version of the dynamics. 
In Section \ref{sec:transversality}, we show that indeed only the surface areas of the transverse parts $\cH_m$ of each $\overline{\cH}_m$ are of interest, since hitting the boundaries $\partial \cH_m$ of the surfaces will be shown to be improbable. 
In Section \ref{sec:Localization}, we make a crucial step in 
the proof of Theorem  \hyperlink{thmA}{A} by contradiction: 
if the set of wandering transition points had positive measure, then in one 
Poincar\'e surface we would find a compact subset, whose intersection with the set 
of these points had positive area. 
The main tool in doing so is a version of the Flow-Box Theorem, which we will state at the beginning of that section. 
Then, in Section~\ref{sec:Transit} we will take a look at the progression of those transition points by defining Poincar\'e maps between the surfaces (hence the name Poincar\'e surfaces), which are shown to be area preserving. 
Since the total area of the surfaces is assumed to decrease to zero, this contradicts an initial set to have positive area.

In Section \ref{sec:Kaehler}, we show how the result can 
be restated if the symplectic manifold is K\"ahler.
Finally, in Section \ref{sec:Applications} we indicate how we apply the
scheme given here in forthcoming articles.
\section{Discretization of the Problem}\label{sec:Disc}
Here we consider a discrete dynamics.
This will be used in Section \ref{sec:Transit} to model the dynamics on the Poincar\'e surfaces.

Let $T:M\to M$ be a continuous injective map of a topological space, 
preserving a Borel measure $\mu:{\cal M}\to [0,\infty]$ on the 
Borel $\sigma$-algebra ${\cal M}$.

The  {\em wandering set} 
\[\Wand\equiv \Wand_T\subseteq M\] 
of the corresponding discrete dynamical system
consists of those $x\in M$ which have a neighborhood  $U_x$ so that 
$U_x\cap T^t(U_x)=\emptyset\quad (t\in\bN)$. 

We denote by $\cO^+(x):=\{T^t(x)\mid t\in \bN_0\}$ the forward orbit of $x\in M$. 
The subsets 
\[{\cH}_m\equiv {\cH}^T_m\in {\cal M}\qquad(m\in\bN)\] 
are assumed to have finite measures.
The {\em transition set} is defined by
\begin{equation}
\label{trans:t}
\Trans\equiv \Trans_T:=\{x\in M\mid\exists\, m_0\in\bN\;\forall \,m\ge m_0: \cO^+(x)\cap {\cH}^T_m\neq\emptyset \}.
\end{equation}

\begin{Lemma}  \label{lem:Disk}
Assuming $\lim_{m\to\infty}\mu({\cH}_m)=0$, then $\mu(\Trans\cap\Wand)=0$.
\end{Lemma}
\textbf{Proof:}
Assuming the converse, we find 
$k\in \bN$ and $x\in {\cH}_{k}\cap \Trans\cap\Wand$
such that $\mu(U_x\cap {\cH}_{k}\cap \Trans\cap\Wand)>0$ 
for all neighborhoods $U_x$ of $x$. 
As $x\in \Wand$, there exists such a neighborhood $U_x\subseteq {\cH}_{k}$
with $U_x\cap T^t(U_x) = \emptyset\quad (t\in\bN)$, so that by injectivity
\begin{equation}
\label{eq:t1:t2}
T^{t_1}(U_x)\cap T^{t_2}(U_x)=\emptyset \qquad (t_1\neq t_2\in\bN_0).
\end{equation}
For such a $k$ and $U_x$ we set ${\cal K}_{k}:=U_x\cap \Trans\cap\Wand$ and
\[{\cal K}_{k,\ell}:= \{y\in{\cal K}_{k} \mid 
|\cO^+(y)\cap{\cH}_{k} |=\ell\} \qquad(\ell\in \bN\cup\{\infty\}).\]
As we have the disjoint union
\[{\cal K}_{k}=\coprod_{\ell\in \bN\cup\{\infty\}}{\cal K}_{k,\ell}\,,\]
by our assumption $\mu({\cal K}_{k})>0$ there is an $\ell$ with $\mu({\cal K}_{k,\ell})>0$.
However:
\begin{enumerate}[$\bullet$]
\item 
We notice that $\mu({\cal K}_{k,\infty})=0$, since otherwise by \eqref{eq:t1:t2}
\[\mu\big(\bigcup_{t\in\bN_0}T^t({\cal K}_{k,\infty})\cap \cH_k\big)
=\sum_{t\in \bN_0} \mu\big(T^t({\cal K}_{k,\infty})\cap \cH_k\big)=\infty,\]
contradicting $\mu({\cH}_{k})<\infty$.
\item 
But also $\mu({\cal K}_{k,\ell})=0$ for all $\ell\in\bN$. For otherwise, using
${\cal K}_{k,\ell}\subseteq\Trans$,  with
\[{\cal K}_{k,\ell,m_0}:=
\{y\in {\cal K}_{k,\ell}\mid \forall \,m\ge m_0: \cO^+(y)\cap {\cH}_m\neq\emptyset\} \qquad(m_0>k)\]  
there exists a $m_0>k$ with $\mu({\cal K}_{k,\ell,m_0})>0$. But this is impossible:
Choose $m\ge m_0$ so that $\mu({\cH}_m)<\mu({\cal K}_{k,\ell,m_0})$,
and let 
\[\tau(y):=\min \{t\in \bN\mid T^t(y)\in {\cH}_m\}\qquad (y\in {\cal K}_{k,\ell,m_0}).\]
Then the map
\[{\cH}_{k,\ell,m_0}\to {\cH}_m\quad,\quad y\mapsto T^{\tau(y)}(y)\]
is one to one, and thus $\mu({\cH}_m)\ge \mu({\cH}_{k,\ell,m_0})$, as $T$ preserves $\mu$.
\end{enumerate}
So we derived a contradicting to the assumption $\mu(\Trans\cap\Wand)>0$.
\hfill$\Box$\\[2mm]
In Section \ref{sec:transversality} we will apply Lemma \ref{lem:Disk}
to a Poincar\'e map for the flow $\Phi$, 
on a certain subset of $\cup_{m\in\bN}\cH_m$.
\section{Transversality}
\label{sec:transversality}
\begin{figure}\centering
\begin{scaletikzpicturetowidth}{0.7\textwidth}
\begin{tikzpicture}[scale=\tikzscale,y=0.5cm]
  \draw[dotted,very thick,->] (-3,0) -- (3,0) node[right] {$\bR$};
  \draw[dotted,very thick,->] (0,-1) -- (0,5) node[above] {$\bR$};
  \draw[domain=-3:3,smooth,variable=\x,very thick,blue] plot ({\x},{\x*\x/2+1}) node[right] {$\cH$};

\draw[green,thick,->] (-3,4.5) -- (3,4.5);
\draw[green,thick,->] (-3,3.5) -- (3,3.5) node[right] {$X_H$};
\draw[green,thick,->] (-3,2.5) -- (3,2.5);
\draw[green,thick,->] (-3,1.5) -- (3,1.5);
\draw[red,very thick] (-3,1) -- (3,1) node[right] {$\cT$};
\draw[green,thick,->] (-3,0.5) -- (3,0.5);
\draw[green,thick,->] (-3,-0.5) -- (3,-0.5);
\end{tikzpicture}
\end{scaletikzpicturetowidth}
 \caption[Transverse and tangent orbits]{Transverse and tangent orbits: here $P = \bR^2$, $X_H = e_1$ and $\cH = \operatorname{Graph}(x \mapsto x^2 +1)$, so the flow is
 transverse to $\cH \setminus \{(0,1)\}$ and $\cT = \bR \times \{1\}$.}
 \label{fig:Transverse}
\end{figure}
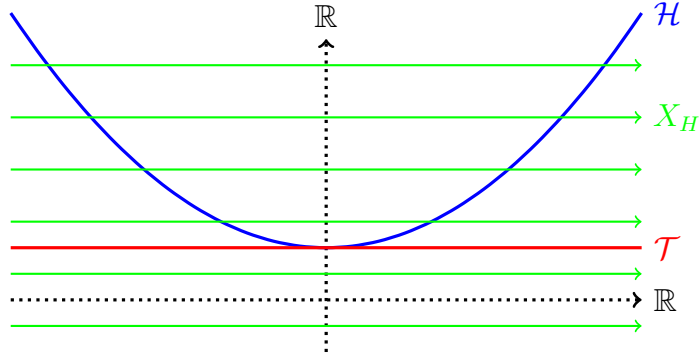

In the following sections, we need to move points along their orbits, until they hit one of the Poincar\'e surfaces. We usually need the orbits to be transverse to the surfaces
to ensure that the corresponding (local) Poincar\'e  maps are $C^1$. 
Also, we want to define a global {\em continuous} discrete dynamics for using 
Lemma \ref{lem:Disk}.
As we would like to exclude orbits hitting a boundary $\partial\cH_m$ 
from our considerations early on, we claim:

\begin{Lemma}
\label{lem:TangentPoints}
The measure $\Omega(\cT)$ vanishes for the set 
\[
\cT:= \bigcup_{m\in\bN} \cT_m\qquad\mbox{with}\quad 
\cT_m := \left\{ x  \in P \mid \cO(x) \cap \partial\cH_m \neq\emptyset\right\}
\quad (m\in\bN).
\]
\end{Lemma}
\begin{Remark}[Transversality]\quad\\[-6mm] \label{rem:trans:Sard}
\begin{enumerate}[1.]
\item 
Hence, if we want to prove Theorem \hyperlink{thmA}{A}, {\em i.e.} show that 
$\Omega(\Trans\cap\Wand) = 0$, we only need to show  
$\Omega(\Trans\cap\Wand \cap \cT^C) = 0$, since $\cT$ 
is of measure zero.
\item 
$\cT$ contains all phase space points $x$, whose orbit 
$\cO(x)$ hits the submanifolds $\overline{\cH}_m$ tangentially.
Applying Sard's Theorem, we could even show the analog of
Lemma \ref{lem:TangentPoints} if we allowed the flow to be tangential to 
the relative interior $\,\cH_m$ of $\,\overline\cH_m$,
see Figure \ref{fig:Transverse}. \hfill $\Diamond$
\end{enumerate}
\end{Remark}
{\bf Proof of Lemma \ref{lem:TangentPoints}:}
We must show that $\Omega(\cT_m) = 0$ for all $m\in \bN$.
But as $\partial\cH_m$ is a codimension two embedded submanifold of $P$
and $\Phi\in C^1(D,P)$,
\[\cT_m=\Phi\big(D\cap(\bR\times\partial\cH_m)\big) \]
is of measure zero.
\hfill $\Box$\\[2mm]
{\bf Proof of Theorem \hyperlink{thmA}{A}:}\\
We now use Lemma \ref{lem:Disk} in the following way.
\begin{enumerate}[$\bullet$]
\item 
$M:=\cup_{m\in \bN}{\cH}^T_m\quad$ with $\quad{\cH}^T_m:={\cH}_m\cap \Set$,\\ 
is the union of the parts of the Poincar\'e surfaces, belonging to  
\[\Set:=\Trans\cap \Wand\setminus\cT.\]
As $\Set\cap \cT=\emptyset$,  the flow $\Phi$ is transversal to $M$.
\item 
$T:M\to M$ is defined as the next intersection with a Poincar\'e surface.
As $\Set\subseteq \Trans$, this exists. So the {\em return time}
\[\tau:M\to\bN\cup\{\infty\}\quad, \quad \tau(x)=\inf\{t\in (0,T(x))\mid \Phi(t,x)\in M\} \]
is finite. Since $\Phi$ is transversal to $M$, it is positive, and one sets
\[T(x):=\Phi(\tau(x),x).\]
Notice that by transversality $T\in C(M,M)$. By definition $T$ is injective.
\item 
$\mu:=\imath_M^* {\cal V}$ is the invariant  measure, with $\imath_M:M\to P$. 
So $\mu({\cH}^T_m)={\cal V}({\cH}_m)<\infty$, using Lemma \ref{lem:TangentPoints}.
In particular, the assumption $\lim_{m\to\infty}\mu({\cH}_m)=0$ of Lemma  \ref{lem:Disk} 
is true.

Moreover, by Lemma \ref{lem:PoincareMap} the map $T$ preserves the measure $\mu$.
\end{enumerate}
So in this application the whole topological space $M$ is now the wandering set of the
continuous transformation $T$ and also equals $\Trans_T$ as defined in \eqref{trans:t}.
We apply Lemma \ref{lem:Disk} to show that $\mu(M)=0$.
By Lemma \ref{lem:LocalVol} below this implies that  $\Omega(\Trans\cap \Wand) = 0$.
\hfill$\Box$

\section{Localization to one Poincar\'e Surface}
\label{sec:Localization}
As pointed out in the introduction, one step of proving Theorem \hyperlink{thmA}{A} is 
the following implication: \\
if $\Trans_\Phi$ had positive measure, then there would be some compact subset of 
some Poincar\'e surface, whose intersection with $\Trans_\Phi$ had positive area:

\begin{Lemma}
\label{lem:LocalVol}
Suppose $\Omega(\Trans\cap \Wand) > 0$.
Then there exist $m\in\bN$ and a compact subset $\cK_{m} \subseteq \cH_{m}$ with
\begin{equation*}
\cV_{m} \left(\cK_{m} \cap \Set \right) > 0.
\end{equation*}
\end{Lemma}

Our main tool in proving this is the following Flow-Box Theorem, which identifies the change of volume of sets like $\Phi( (-t,t) \times \cK )$ (with an appropriate codimension one 
$\partial$-submanifold $\cK$) in ongoing time with the surface area of the submanifold $\cK$.
To be more precise:

\begin{Lemma}[Flow-Box Theorem]
\label{lem:FlowBox}
Let $\cH\subseteq P$ be a submanifold of codimension one, 
such that the vector field $X$ is transverse to $\cH$. Then $x\in \cH$ has
a compact $\partial$-submanifold $\cK\subseteq \cH$ as a neighborhood, 
$L_t := \Phi( (-t,t) \times \cK )$ is a $\partial$-submanifold of $P$ for sufficiently small $t>0$, and
\begin{equation}
 \label{eq:FloxBox}
 \int_{L_t} F\, \Omega  = 2t \int_{\cK} f\,{\cal V}
\end{equation}
for all continuous functions $f$ on $\cK$;
here, $F$ is the continuation of $f$ to $L_t$ that is constant along the flow lines, i.e.
\begin{equation}
\label{F}
F: L_t \rightarrow \bC, \quad F(\Phi(s,x)) := f(x) \qquad (s\in(-t,t),x\in \cK).
\end{equation}
\end{Lemma}

\begin{figure}\centering 
\begin{scaletikzpicturetowidth}{0.9\textwidth}
\begin{tikzpicture}[scale=\tikzscale]
  \coordinate[label=below:$P$] (P) at (0,0);  
  
  \draw (P) to [out=90,in=-90](3,6) to [out=180,in=0](-3,7) to [out=-90,in=90](-6,1) to [out=0,in=180](P);
  
  \coordinate (L0) at (-1,1);
  \coordinate[label=above:\textcolor{red}{$L_\varepsilon$}] (L1) at (1.5,5);
  
  \draw[name path=right,red,thick,dashed] (L0) to [out=90,in=-90](L1);
  \draw[red,thick,<-] (L1) to [out=180,in=0](-2.5,6);
  \draw[red,thick,dashed] (-2.5,6) to [out=-90,in=90](-5,2);
  \draw[red,thick,->] (-5,2) to [out=0,in=180](L0);
  
  \coordinate[label=below:\textcolor{blue}{$\cK$}] (H) at (-3,1.5);
  
  \draw[blue,thick] (H) to [out=90,in=-90]++(2.5,4);
  
  \path [name path=height2] (-6,2) -- (3,2);
  \path [name path=height3] (-6,3) -- (3,3);
  \path [name path=height4] (-6,4) -- (3,4);
  
  \path [name intersections={of = height2 and right}];
  \coordinate (R2)  at (intersection-1);
  \path [name intersections={of = height3 and right}];
  \coordinate (R3)  at (intersection-1);
  \path [name intersections={of = height4 and right}];
  \coordinate (R4)  at (intersection-1);
  
  \draw[red,thick,<-] (R2) to [out=180,in=0]++(-4,1);
  \draw[red,thick,<-] (R3) to [out=180,in=0]++(-4,1);
  \draw[red,thick,<-] (R4) to [out=180,in=0]++(-4,1);
  
  \draw[->] (4,3) to [out=30,in=180] (5,3.3) node [above] {$\varphi$} to [out=0,in=150](6,3);
  
  \draw[->] (8,0) to (14,0) node [below right] {$\bR$};
  \draw[->] (11,-0.5) to (11,6) node [above] {$\bR^{2n-1}$};
  
  \draw[blue,thick] (11,1) node [right] {\textcolor{blue}{$W$}} to (11,5);
  
  \draw (9,-0.2) node [below] {$-\varepsilon$} to (9,0.2);
  \draw (13,-0.2) node [below] {$\varepsilon$} to (13,0.2);
  
  \draw[red,thick,->] (9,1)--(13,1);
  \draw[red,thick,->] (9,2)--(13,2);
  \draw[red,thick,->] (9,3) to (11,3) to (13,3);
  \draw[red,thick,->] (9,4)--(13,4);
  \draw[red,thick,->] (9,5)--(13,5);
  
  \draw[red,thick,dashed] (9,1)--(9,5);
  \draw[red,thick,dashed] (13,1)--(13,5);
\end{tikzpicture}
\end{scaletikzpicturetowidth}
  \caption[Flow-box chart]{Flow-box chart $\varphi:L_\varepsilon\rightarrow (-\varepsilon,\varepsilon) \times W$ with $\cK = \varphi^{-1}(\{0\}\times W)$.}
  \label{fig:FlowBox}
\end{figure}
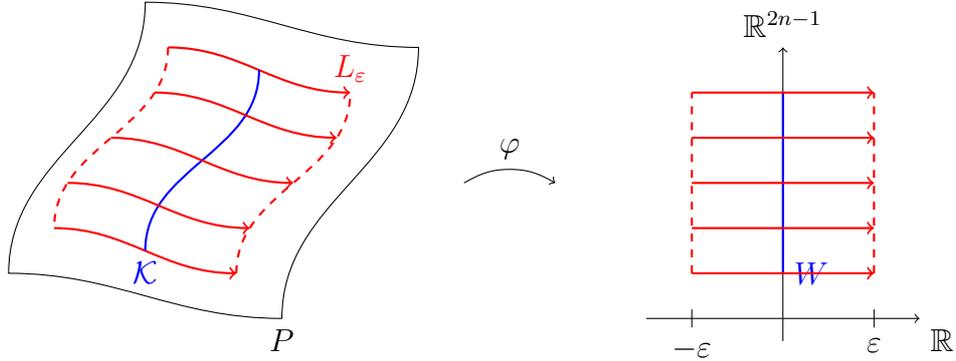

{\bf Proof:}
Let $x\in \cH$. As initially assumed (Remark \ref{rem:wand}.3), $X(x)\neq0$.
By the 'straightening out' Theorem 2.1.9 in \cite{Abraham_Marsden_1978}, this means 
that on a suitable open neighborhood $U$ of $x$ in $P$
there exists a chart $\varphi:U \rightarrow I \times W \subseteq \bR \times \bR^{2n-1}$, 
where
$I:=(-\varepsilon,\varepsilon)$ for $\varepsilon>0$, 
and $W \subseteq \bR^{2n-1}$ is open, such that (see Figure \ref{fig:FlowBox})
\begin{enumerate}
\item 
$I \ni \lambda \mapsto \varphi^{-1}(\lambda,\eta)$ is an integral curve of $X$ for all 
$\eta \in W$.
\item 
$\varphi^{-1}(\{0\}\times W) = \cH \cap U$.
\item 
The local representative of the vector field $X$ at any point of the chart is 
$e_1=(1,0,\ldots,0)\in\bR^n$.
\end{enumerate}

We choose a  $\partial$-submanifold $\cK\subseteq \cH\cap U$ as a compact neighborhood of $x$.
Then for $t\in (0,\varepsilon)$ 
\[\varphi(L_t)=(-t,t)\times \varphi(\cK).\]
So $L_t$ is a $\partial$-submanifold of $P$, and its closure $\overline{L_t}$
is a submanifold with corners.
By Property 3.\ we can write the volume form $\Omega$ on $L_\varepsilon$ as
\[\Omega=\bsd t\wedge {\cal V} ,\]
using the time coordinate $t$ of the chart $\varphi$.
Therefore 
\begin{equation}
\label{FOmega}
F\,\Omega=\bsd t\wedge F\,{\cal V}=\bsd t\wedge \bsi_XF\Omega.
\end{equation}
As the Lie derivative ${\cal L}_XF\Omega= ({\cal L}_XF)\Omega +F{\cal L}_X\Omega$ vanishes
by Def.\ \eqref{F}, we have $\bsd F{\cal V}=\bsd\bsi_{X}F\Omega=0$, too.
So \eqref{FOmega} implies $F\,\Omega=\bsd(t \bsi_XF\Omega)= \bsd(t F{\cal V})$, and Stokes' Theorem tells us that
\[ \int_{\overline{L_t}} F\Omega=\int_{\overline{L_t}}  \bsd(t F\,{\cal V})
=\int_{\partial\overline{L_t}}tF\,{\cal V}=2t \int_{\cK}f\,{\cal V}.\]
In the last equation we used that ${\cal V}= \bsi_X\Omega$ 
vanishes on $\Phi((-t,t),\partial \cK)$.
\hfill$\Box$\\[2mm]

\begin{figure}\centering 
\begin{scaletikzpicturetowidth}{0.8\textwidth}
\begin{tikzpicture}[scale=\tikzscale]
 \draw[->, name path=Flow,thick,red] (0,0) to [out=0,in=180] (22,1) node[right] {\textcolor{red}{$X_H$}};
 
 \draw[dashed, name path=left,red] (1,-2.5) to [out=90,in=-90] (2,2.5);
 \draw[dashed,red] (5,-2.4) to [out=90,in=-90] (6,2.6);
 
  \path [name path=height2] (0,-1.5) -- (20,-1.5);
  \path [name path=height3] (0,-0.5) -- (20,-0.5);
  \path [name path=height4] (0,0.5) -- (20,0.5);
  \path [name path=height5] (0,1.5) -- (20,1.5);
  \path [name path=height6] (0,2.5) -- (20,2.5);
  \path [name path=height7] (0,3.5) -- (20,3.5);
  
  \path [name intersections={of = height2 and left}];
  \coordinate (L2)  at (intersection-1);
  \path [name intersections={of = height3 and left}];
  \coordinate (L3)  at (intersection-1);
  \path [name intersections={of = height4 and left}];
  \coordinate (L4)  at (intersection-1);
  \path [name intersections={of = height5 and left}];
  \coordinate (L5)  at (intersection-1);
  
  \draw[dashed,->,red] (1,-2.5) to [out=0,in=180]++(4,0.1);
  \draw[dashed,->,red] (2,2.5) to [out=0,in=180]++(4,0.1) node[above] {\textcolor{red}{$\Phi_{-t_{m}}(L_\varepsilon)$}};
  \draw[dashed,->,red] (L2) to [out=0,in=180]++(4,0.1);
  \draw[dashed,->,red] (L3) to [out=0,in=180]++(4,0.1);
  \draw[dashed,->,red] (L4) to [out=0,in=180]++(4,0.1);
  \draw[dashed,->,red] (L5) to [out=0,in=180]++(4,0.1);

  \draw[dashed, name path=right,red] (14,-1.5) to [out=90,in=-90] (15,3.5);
  \draw[dashed,red] (18,-1.4) to [out=90,in=-90] (19,3.6);
 
  \path [name intersections={of = height3 and right}];
  \coordinate (R2)  at (intersection-1);
  \path [name intersections={of = height4 and right}];
  \coordinate (R3)  at (intersection-1);
  \path [name intersections={of = height5 and right}];
  \coordinate (R4)  at (intersection-1);
  \path [name intersections={of = height6 and right}];
  \coordinate (R5)  at (intersection-1);
  
  \draw[dashed, name path=Lbot,->,red] (14,-1.5) to [out=0,in=180]++(4,0.1);
  \draw[dashed, name path=Ltop,->,red] (15,3.5) to [out=0,in=180]++(4,0.1) node[above] {\textcolor{red}{$L_\varepsilon$}};
  \draw[dashed,->,red] (R2) to [out=0,in=180]++(4,0.1);
  \draw[dashed,->,red] (R3) to [out=0,in=180]++(4,0.1);
  \draw[dashed,->,red] (R4) to [out=0,in=180]++(4,0.1);
  \draw[dashed,->,red] (R5) to [out=0,in=180]++(4,0.1);
  
  \path[name path=H] (17,6) to [out=-90,in=90] (16,-5);
  
  \path [name intersections={of = Flow and H}];
  \coordinate[label=below right:$x_{m}$] (Xm)  at (intersection-1);
  
  \path [name intersections={of = Ltop and H}];
  \coordinate (Ktop)  at (intersection-1);
  \path [name intersections={of = Lbot and H}];
  \coordinate[label=right:\textcolor{blue}{$\cK_{m}$}] (Kbot)  at (intersection-1);
  
  \draw[very thick,dotted,blue] (Ktop) to [out=-90,in=90] (Kbot);
  \draw[thick,blue] (Ktop) to ++(0,3);
  \draw[thick,blue] (Kbot) to ++(0,-3) node[above right] {\textcolor{blue}{$\cH_{m}$}};
  
  \draw[dotted,black] (Xm) circle (5);
  \coordinate[label=right:\textcolor{black}{$B_\delta(x_{m})$}] (B) at (18,6);
  
  \path[name path=Hl] (4,5) to [out=-90,in=90] (3,-6);
  \path [name intersections={of = Flow and Hl}];
  \coordinate[label=below right:$x_0$] (X0)  at (intersection-1);
  
  \foreach \p in {Xm,X0} \fill [black] (\p) circle (3.0pt);
  
  \path[name path=Hm] (10,5.5) to [out=-90,in=90] (9,-5.5);
  \path [name intersections={of = Flow and Hm}];
  \coordinate[label=below right:\textcolor{red}{$t_{m}$}] (T)  at (intersection-1);
\end{tikzpicture}
\end{scaletikzpicturetowidth}
  \caption[Localization on a Poincar\'e surface]{Localization on a Poincar\'e surface: we illustrate the points $x_0$ and $x_{m} \in \cH_{m}$ as constructed in the proof, as well as the sets $B_\delta(x_{m})$ and $\cK_{m}$ plus $L_\varepsilon$ and $\Phi_{-t_{m}}(L_\varepsilon)$.}
  \label{fig:Localization}
\end{figure}
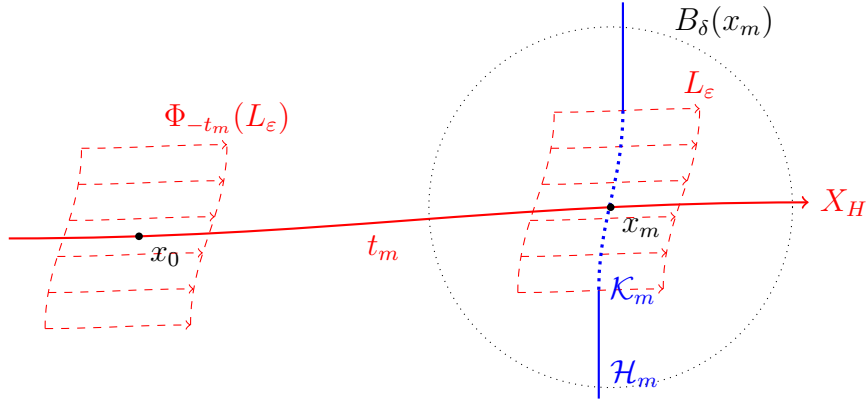

With this, we can give the proof of the preceding lemma:\\[2mm]
{\bf Proof of Lemma \ref{lem:LocalVol}:}
To simplify notation, we use the flow invariant set
\[ \Set=\Trans\cap \Wand\setminus\cT.\]
By Lemma \ref{lem:TangentPoints}, $\Omega(\Trans\cap \Wand) > 0$ 
implies $\Omega\left(\Set \right) > 0$ as well.
In this case, there exists $x_0\in \Set$ with
\begin{equation}
\label{eq:locPoint}
 \Omega \left(U\cap \Set\right) > 0 \qquad \mbox{for every neighborhood $U\subseteq  P$ of $x_0$.}
\end{equation}
By definition of the set $\Trans$, for a sufficiently large $m\in\bN$ there exists a time $t_{m} \in (0,T^+(x_0))$ with 
\begin{equation*}
x_{m}:= \Phi(t_{m},x_0) \in \overline{\cH}_{m}.
\end{equation*}
Specifically because of $x_0 \in \cT^C$, we have that $x_{m} \in \cH_{m}$.
For a sufficiently small $\delta > 0$, 
we have that\,\footnote{with respect to an arbitrary Riemannian metric $g$ on $P$ and the metric $d$
induced by $g$} the ball 
$B_\delta(x_{m}):=\{y\in P\mid d(y,x_0)<\varepsilon\}$ is a subset of
\begin{equation*}
 \left\{ x \in  P\mid (-t_{m},x) \in D \right\},
\end{equation*}
since the latter is open (see \cite[Th. 17.8]{Lee_2003}) and contains  $x_{m}$; 
here, $D$ is the domain of the flow $\Phi$, see  \eqref{eq:DefD}.
Then
\begin{equation*}
\cK_{m} := \overline{\cH_{m} \cap B_{\delta/2}(x_{m})}
\end{equation*}
is a compact subset of $\cH_{m}$ with $x_{m} \in \cK_{m}$, which just like $\cH_{m}$ 
is of codimension one in $ P$.
Then we find $\varepsilon > 0$, such that for every point in $\cK_{m}$ the flow 
exists at least on the time interval $(-\varepsilon,\varepsilon)$; then
\begin{equation*}
L_\varepsilon := \Phi\left( (-\varepsilon,\varepsilon) \times \cK_{m}\right)
\end{equation*}
is well-defined, see Figure \ref{fig:Localization}.

By a possible diminishment of $\varepsilon$, we have that
$L_\varepsilon \subseteq B_\delta(x_{m})$,
hence $\Phi_{-t_{m}}(L_\varepsilon)$ is well-defined.
This is a neighborhood of $x_0$, such that this point's construction implies that
\begin{equation*}
\Omega\left(\Phi_{-t_{m}}(L_\varepsilon)\cap \Set \right) > 0,
\end{equation*}
see \eqref{eq:locPoint}.
Thus
\begin{equation}
\label{eq:FlowBoxVol}
 \Omega \left(L_\varepsilon\cap \Set \right) > 0,
\end{equation}
since by flow invariance of $\Set$
\begin{equation*}
\Phi_{-t_{m}}(L_\varepsilon)\cap \Set= \Phi_{-t_{m}}\left(L_\varepsilon\cap \Set \right)
\end{equation*}
and $\Phi_{-t_{m}}$ is volume preserving, see \cite[Prop 3.3.4]{Abraham_Marsden_1978}.

Now we show that $\mu_{m} \left( \cK_{m} \cap \Set \right) > 0$. 
This is an immediate consequence of the Flow-Box Theorem, see Lemma \ref{lem:FlowBox}:
equation \eqref{eq:FloxBox} yields that
\begin{equation}
\label{eq:FloxBoxInt}
 \int_{L_\varepsilon} F \, \Omega = 2\varepsilon \int_{\cK_{m}} f \, \cV
\end{equation}
for all continuous functions $f$ on the compact set $\cK_{m}$, where $\Omega$ is 
the canonical volume form on $P$; for $F$, we adopt the notation from 
Lemma \ref{lem:FlowBox} (continuation of $f$, which is constant along the flow lines).

Now let $f_n$ ($n\in\bN$) be a sequence of smooth functions on $\cH_{m}$ with compact support in $\cK_{m}$, with 
$L^1\!-\!\lim_{n\rightarrow \infty} f_n = \1_{\cK_{m} \cap \Set}$. 
Thus
\begin{equation}
\label{eq:FlowBoxApprox}
 2 \varepsilon \int_{\cH_{m}} f_n \, \cV_m 
\stackrel{\eqref{eq:FloxBoxInt}}{=} \int_{L_\varepsilon} F_n \, \Omega \qquad(n\in\bN)
\end{equation}
where again $F_n$ is the continuation of $f_n$, that is constant along the flow lines. 
Hence, $L^1\!-\!\lim_{n\rightarrow \infty} F_n = \1_{L_\varepsilon \cap \Set}$, 
so taking the limit 
$n\rightarrow\infty$ in \eqref{eq:FlowBoxApprox} exactly yields
\begin{equation*}
2\varepsilon \cdot \cV_{m} \left(\cK_{m} \cap \Set \right) = 
\Omega\left(L_\varepsilon \cap \Set \right).
\end{equation*}
By \eqref{eq:FlowBoxVol}, it follows that $\cV_{m} \left(\cK_{m} \cap \Set \right)>0$.
\hfill $\Box$

\section{Transit between the Poincar\'e Surfaces}
\label{sec:Transit}

Below, we will identify points on one surface with certain elements on their positive semi-orbit, lying on another surface.
We will call such a mapping \emph{Poincar\'e map}, and 
show that they are 
volume preserving.

Now we give the general notion of Poincar\'e maps as used within this paper.
We first consider general (incomplete) flows and then specialize to the Hamiltonian
flow $\Phi$, and to its restriction to energy surfaces.

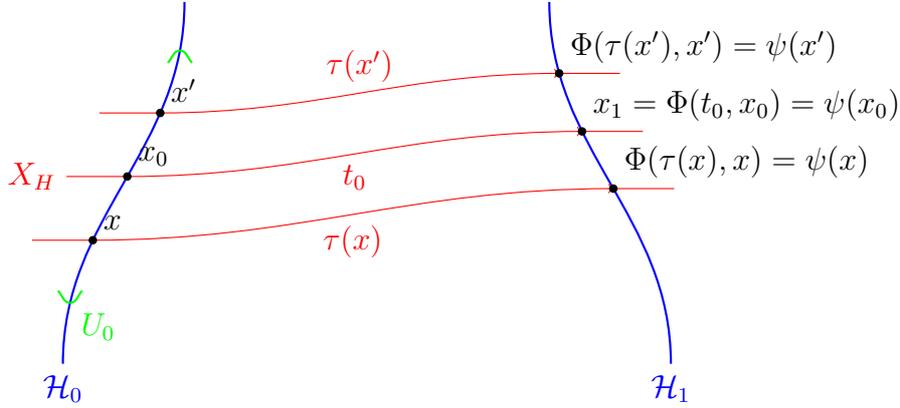
\begin{figure}\centering 
\begin{scaletikzpicturetowidth}{0.8\textwidth}
\begin{tikzpicture}[scale=\tikzscale]
 \draw[name path=H0,thick,blue] (0,0) node [below] {\textcolor{blue}{$\cH_0$}} to [out=90,in=-90](2,6);
 \draw[name path=H1,thick,blue] (10,0) node [below] {\textcolor{blue}{$\cH_1$}} to [out=90,in=-90](8,6);
 
 \path [name path=slope1] (0,1) -- (10,2);
 \path [name path=slope2] (0,2) -- (10,3);
 \path [name path=slope3] (0,3) -- (10,4);
 \path [name path=slope4] (0,4) -- (10,5);
 \path [name path=slope5] (0,5) -- (10,6);
 
 \path [name intersections={of = slope1 and H0}];
 \coordinate[label=below right:{\textcolor{green}{$U_0$}}] (U0) at (intersection-1);
 
 \draw[green,thick] (U0) to [out=180,in=-45]++(-0.2,0.2);
 \draw[green,thick] (U0) to [out=0,in=-135]++(0.2,0.2);
 
 \path [name intersections={of = slope5 and H0}];
 \coordinate (U0a) at (intersection-1);
 
 \draw[green,thick] (U0a) to [out=180,in=45]++(-0.2,-0.2);
 \draw[green,thick] (U0a) to [out=0,in=135]++(0.2,-0.2);
 
 \path [name intersections={of = slope2 and H0}];
 \coordinate[label=above right:$x$] (L2) at (intersection-1);
 \path [name intersections={of = slope3 and H0}];
 \coordinate[label=above right:$x_0$] (L3) at (intersection-1);
 \path [name intersections={of = slope4 and H0}];
 \coordinate[label=above right:$x'$] (L4) at (intersection-1);
 \path [name intersections={of = slope2 and H1}];
 \coordinate[label=above right:{$\Phi(\tau(x),x)=\psi(x)$}] (R2) at (intersection-1);
 \path [name intersections={of = slope3 and H1}];
 \coordinate[label=above right:{$x_1=\Phi(t_0,x_0)=\psi(x_0)$}] (R3) at (intersection-1);
 \path [name intersections={of = slope4 and H1}];
 \coordinate[label=above right:{$\Phi(\tau(x'),x')=\psi(x')$}] (R4) at (intersection-1);
 
 \path[red] (L2) edge[out=0,in=180,->] node [below] {\textcolor{red}{$\tau(x)$}} (R2);
 \path[red] (L3) edge[out=0,in=180,->] node [below] {\textcolor{red}{$t_0$}} (R3);
 \path[red] (L4) edge[out=0,in=180,->] node [above] {\textcolor{red}{$\tau(x')$}} (R4);
 
 \draw[red] (L2) to ++(-1,-0);
 \draw[red] (R2) to ++(1,0);
 \draw[red] (L3) to ++(-1,-0) node [left] {\textcolor{red}{$X_H$}};
 \draw[red] (R3) to ++(1,0);
 \draw[red] (L4) to ++(-1,-0);
 \draw[red] (R4) to ++(1,0);
 
 \foreach \p in {L2,L3,L4,R2,R3,R4} \fill [black] (\p) circle (2.0pt);
\end{tikzpicture}
\end{scaletikzpicturetowidth}
  \caption{Poincar\'e map}
  \label{fig:PoincMap}
\end{figure}

\begin{Lemma}[Poincar\'e Maps]\quad\\[-6mm]
\label{lem:PoincareMap}
\begin{enumerate}[1.]
\item 
Let $X$ be a vector field on a manifold $M$ whose flow $\Phi\in C^1(D,M)$
has maximal  domain $D\subseteq \bR\times M$. 
Consider two codimension one submanifolds $\cH_0, \cH_1\subseteq M$, 
such that $X$ is transverse to $\cH_i$ ($i=0,1$).
Fix a point $x_0 \in \cH_0$ and assume there is a time $t_0 > 0$, such that $x_1 := \Phi(t_0,x_0) \in \cH_1$.

Then there exist a neighborhood $U_0\subseteq \cH_0$ of $x_0$ and a 
so-called \emph{hitting time}  $\tau\in C^1(U_0,\bR)$ with 
$\tau(x_0) = t_0$, such that on $U_0$ the \emph{Poincar\'e map} $\psi$,
\begin{equation*}
\psi(x):= \Phi(\tau(x),x) \in \cH_1 \qquad (x\in U_0)
\end{equation*}
is a diffeomorphism onto its image $U_1$.
\item 
Now consider the case of the flow $\Phi$ on the manifold $P$
generated by $X$ and preserving the volume form $\Omega$.
Then the restrictions of ${\cal V}= \bsi_{X}\Omega$ to
$\cH_i$ are volume forms, 
and the \emph{Poincar\'e map},
see Figure \ref{fig:PoincMap}, is volume preserving.
\item 
Finally consider the restriction of a Hamiltonian flow 
$\Phi$ to an energy surface $\Sigma_E$. Then
$\cH_i \subseteq \Sigma_E$ are symplectic submanifolds of $(P,\omega)$, 
and the \emph{Poincar\'e map}
is a symplectomorphism onto its image.
\end{enumerate}
\end{Lemma}

\textbf{Proof:}\\[-6mm]
\begin{enumerate}[1.]
\item 
The proof is analogous to \cite[Theorem 7.1.2]{Abraham_Marsden_1978}, using the
lower continuity of the escape time.
\item 
Consider for the unit ball $B\subseteq \bR^{2(n-1)}$
a (small) embedded $\partial$-submanifold $\imath_0:B\to V_0\subseteq U_0$ 
with $\imath_0(0)=x_0$ and its image
under the Poincar\'e map $\psi$, the $\partial$-submanifold $\imath_1:V_1:=\psi(V_0)\to U_1$.
We extend these embeddings to the embedding 
\[\imath:[0,1]\times B\to P\quad,\quad \imath(t,x)=\Phi(t\tau(x),x)\]
of a cylinder. 
Then $\int_{[0,1]\times B}\imath^*\bsd{\cal V}=0$, since ${\cal V}$ is closed:
\[\bsd{\cal V}=\bsd\bsi_X \Omega=(\bsd\bsi_X+\bsi_X\bsd)\Omega={\cal L}_X\Omega=0.\]
Similarly $\int_{[0,1]\times \partial B}\imath^*{\cal V}=0$, since $X$ is tangential to
the $\partial$-submanifold $\imath([0,1]\times \partial B)$. Thus by Stokes' theorem
for manifolds with corners 
\[0=\int_{[0,1]\times B}\hspace*{-2mm}\imath^*\bsd{\cal V} 
= \int_{\partial ([0,1]\times B)}\hspace*{-2mm}\imath^*{\cal V}
= \int_{\{1\}\times B}\hspace*{-2mm} \imath^*{\cal V} - \int_{\{0\}\times B}\hspace*{-2mm} \imath^*{\cal V} 
= \int_{B} \imath^*_1{\cal V} - \int_{B} \imath^*_0{\cal V} .\]
\item 
The proof goes along the lines of \cite[Lemma 8.2]{McDuff_Salamon_1995}.
\hfill$\Box$
\end{enumerate}

\section{Submanifolds of K\"ahler Manifolds}
\label{sec:Kaehler}

If there is an underlying metric on the manifold inducing the symplectic structure, then it may be more convenient from a technical point of view, to use the metric structure 
to estimate the symplectic volume of a codimension two
submanifold instead of directly integrating the volume form 
$\bsi_{X_H} \sigma$. In this section, we show that this is possible.

Let $(P,g,J)$ be a $2n$-dimensional K\"ahler manifold, with Riemannian metric $g$ and complex structure $J$, which induce a symplectic form $\omega$ by 
\[\omega(X,Y) := g(JX,Y)\qquad (X,Y\in\cX(P)).\] 
Assume\,\footnote{As explained in Remark \ref{rem:wand}.3, 
this is possible without loss of generality.} that 
$E\in\bR$ is a regular value of $H$, such that $\Sigma_E$ is a codimension-1 submanifold (if non-empty) of $P$. Let $(\cH^E_m)_{m\in\bN}$ be a sequence of submanifolds of $\Sigma_E$ of codimension one.

Let $dP$ be the Riemannian volume form on $P$ induced by $g$; then $dP$ equals the canonical volume form $\Omega$ as defined above. Further, let $d\Sigma_E$ and $d\cH^E_m$ be the Riemannian volume forms on $\Sigma_E$ and $\cH^E_m$ respectively, induced by the restrictions $g_{\Sigma_E}$ resp. $g_{\cH^E_m}$ of $g$ to $\Sigma_E$ resp. $\cH^E_m$.

Then it is sufficient to calculate the Riemannian surface areas of the submanifolds, in order to apply the technique from above.

The following two lemmas will be used in the proof of Theorem \hyperlink{thmB}{B}.
\begin{Lemma}  \label{lem:kaehler}
Let $\imath:N\to P$ be a submanifold of codimension two,
being the preimage of a regular value of a smooth map
$F\equiv (F_1,F_2): P\to \mathbb R^2$. Then
\[\frac{\imath^* \omega^{\wedge n-1}}{(n-1)!}
=\frac{ \omega(X_{F_1},X_{F_2})}
{\sqrt{\|\nabla F_1\|^2\|\nabla F_2\|^2-g(\nabla F_1,\nabla F_2)^2}} \,dN,\]
$dN$ being the Riemannian volume form on $N$ induced by $g$.
\end{Lemma}
\noindent
\textbf{Proof:}
We denote the Hodge star operator by $\star:\Omega^{k}(P)\to \Omega^{2n-k}(P)$.
So for $\phi,\psi\in \Omega^{k}(P)$ and the Riemannian volume form $dP$ on $P$
we have $\phi\wedge \star\psi= \langle \phi,\psi\rangle \,dP$, with the bilinear extension
of
\[\langle \alpha_1\wedge\ldots\wedge \alpha_k,\beta_1\wedge\ldots\wedge \beta_k\rangle 
:= \det\big(\langle \alpha_i,\beta_j\rangle_{i,j=1}^k\big)\qquad
(\alpha_i,\beta_j\in \Omega^{1}(P))\]
and $\langle \alpha_i,\beta_j\rangle:=g(\alpha_i^\#,\beta_j^\#)$ 
(with $g(\gamma^\#,X):=\gamma(X)$ for $\gamma \in \Omega^{1}(P)$ 
and all $X\in \mathcal X(P)$).
It is well known (see, e.g.\ \cite{Ba}) that  
$\star\omega^{\wedge n-k}/(n-k)! =\omega^{\wedge k}/k!$. 
The formula follows by applying $\star$ to both sides, since at any $x\in N$
the denominator 
$\sqrt{\|\nabla F_1\|^2\|\nabla F_2\|^2-g(\nabla F_1,\nabla F_2)^2}$ is the 
Riemannian volume in the normal bundle of $N$, 
spanned by $\nabla F_1(x),\nabla F_2(x)\in T^\perp_xN$,
whereas  $\omega(X_{F_1},X_{F_2})(x)= \omega(\nabla F_1(x),\nabla F_2(x))$.
\hfill $\Box$\\[2mm]
So the Riemannian volume form $dN$ is multiplied by
the normalized Poisson bracket $\{F_1,F_2\}$.
\begin{Lemma}[Wirtinger's inequality]  \label{lem:Bessel}\quad\\
On an $\bR$-vector space $V$ with scalar product $\langle\cdot , \cdot \rangle$, 
complex structure $J:V\to V$ and K\"ahlerian symplectic form $\omega$ one has the inequality
\[\omega(X,Y)^2\leq \|X\|^2\|Y\|^2-\langle X,Y\rangle^2\qquad(X,Y\in V).\]
\end{Lemma}
\textbf{Proof:}
We assume without loss of generality that $\|X\|=1$. Then
the formula follows from Bessel's inequality, as $\omega(X,Y)=\langle JX,Y\rangle$, and
the pair $\{X,JX\}\subseteq V$ is an orthonormal system.
\hfill $\Box$\\[2mm]

Now we  use Lemma \ref{lem:kaehler} and \ref{lem:Bessel} to prove Theorem
\hyperlink{thmB}{B} from Section \ref{sec:Intro}, which we cite for convenience:

\begin{TheoremN}\hspace*{-1mm}{\hyperlink{thmB}{\bf B}} \ 
On any energy surface $\Sigma_E\subseteq P$ in a K\"ahler  manifold 
it follows from Assumptions 1.\ and 2'.\ 
for the Hamiltonian flow that $\ \sigma(\Trans_E\cap\Wand_E) = 0$. 
\end{TheoremN}
\textbf{Proof.}
We show $\left.\left(\bsi_{X_H}  \sigma\right)\right|_{\cH^E_m} = \eta\, d\cH^E_m$ with a function $\eta$, that (in modulus) is bounded above by 1. Then the assertion follows from Theorem \hyperlink{thmA}{A}.

Locally, $\cH^E_m$ is the preimage of the regular value $(E,0)$ of a smooth map 
$(H,F):P\rightarrow \bR^2$. 
With $Y:=\|\nabla H\|^{-2}\nabla_H$ we may take
\[\sigma:= \bsi_Y\Omega,\]
since then
$dH\wedge \sigma= dH\wedge \bsi_Y\Omega= (\bsi_YdH)\wedge \Omega=\Omega,$
using $\bsi_{\nabla H}dH=g(\nabla H,\nabla H)=\|\nabla H\|^2$.
With $s:=(-1)^{\lfloor n/2 \rfloor}$ we have
\begin{equation}
\label{eq:sigma:omega}
\bsi_{X_H}  \sigma = - \textstyle{\frac{s}{(n-1)!}}\omega^{\wedge n-1}\,,
\end{equation}
since 
\[\bsi_{X_H}  \sigma=\|\nabla H\|^{-2} \bsi_{X_H}\bsi_{\nabla H}\Omega
=s\|\nabla H\|^{-2} \bsi_{X_H}\bsi_{\nabla H}\frac{\omega^{\wedge n}}{n!}
=s \frac{n}{n!}\;  \frac{\omega(\nabla H,X_H)}{\|\nabla H\|^{2}}\omega^{\wedge n-1}.\]

By applying Lemma \ref{lem:kaehler}, we get 
$\bsi_{X_H}\sigma\rstr_{\cH^E_m} = \eta\, d\cH_m$ with
\begin{equation}
\label{eq:BoundEta}
\eta:= \frac{\omega(\nabla H, \nabla F) }{\sqrt{\|\nabla H\|^2 \|\nabla F\|^2 - g(\nabla H,\nabla F)^2}}.
\end{equation}
Then $|\eta|\leq 1$ follows from Lemma \ref{lem:Bessel}, applied to the tangent spaces 
$T_xP$ for $x\in\cH^E_m$.
\hfill$\Box$\\[2mm]
Finally we consider the  metric $g$ on the cotangent bundle $\pi:T^*M\to M$ of a Riemannian manifold
$(M^n,h)$, given at $x\in T^*M$ by 
\[g_x((Q_1,P_1),(Q_2,P_2)):= h_{\pi(x)}(Q_1,Q_2)+h^{-1}_{\pi(x)}(P_1,P_2)
\qquad((Q_i,P_i)\in T_xT^*M).\]
The tangent bundle $TT^*M$ of the phase space splits as
\[TT^*M ={\rm ver}\oplus {\rm hor}\]
with the vertical subspace ${\rm ver}_x\subseteq T_xT^*M$ given by the kernel of $T_x\pi^*$, and the
horizontal subspace ${\rm hor}_x\subseteq T_xT^*M$ defined by the Riemannian connection.
With the natural vector bundle isomorphisms 
\[I_{\rm ver}: {\rm ver}\to TM\ \mbox{ and }\ I_{\rm hor}: {\rm hor}\to TM\] 
(induced by $T\pi\rstr_{{\rm hor}}:{\rm hor}\to TM$, respectively by the restriction of the connection, 
see, e.g.\ Klingenberg \cite[Proposition 1.5.10]{Kli95}), 
the vector bundle isomorphism $J$ on $TT^*M\to T^*M$, given by 
\[J_x= 
\left(\begin{array}{cc}0& I_{x,{\rm ver}}^{-1}\circ I_{x,{\rm hor}}\\ 
-I_{x,{\rm hor}}^{-1}\circ I_{x,{\rm ver}}&0 \end{array}\right)
\qquad(x\in T^*M).\]
is an almost complex structure.\,\footnote{i.e.,  a smooth (1,1) tensor field with $J^2=-1$.\\
As shown by Dombrowski in \cite[Appendix III]{Do} (for the tangent bundle $TM$ instead of the
cotangent bundle $T^*M$), 
$J$ defines a K\"ahler structure on $T^*M$, if $(M,h)$ is of vanishing curvature.}
So the symplectic manifold $(T^*M,\omega_0)$ with $J$ and $g(u,v)=\omega_0(u,Jv)$ is an almost 
K\"ahler manifold.

For a Hamiltonian $H\in C^2(T^*M,\bR)$ of the form
\begin{equation}
\label{Ham}
H(q,p)=\eh h^{-1}_{q}(p,p)+ V(q),
\end{equation}
a regular value $E\in V(\bR)$ of $V$ and a hypersurface $\cF\subseteq M$ of finite Riemannian volume, 
we obtain a  hypersurface in $\Sigma_E=H^{-1}(E)$ given by
\[\cH_E:=\{(q,p)\in \Sigma_E\mid q\in\cF\} = \{(q,p)\in T^*_\cF M \mid h_q^{-1}(p,p)=2(E-V(q))\}. \]
So $\cH_E$ is of codimension two in $T^*M$.  By localization, if necessary we can 
assume that $\cF$ is orientable and denote by $N:\cF\to T_\cF M$ a (continuous) unit normal vector 
field. Then (applying $p\in T^*_qM$ to $N\in T_qM$)
\[\cH_E^\pm := \{(q,p)\in \cH_E\mid \pm\, p(N(q))>0\}\]
both project diffeomorphically to their common image in $T^*\cF\subseteq T^*M$, via
\[n^\pm: \cH_E^\pm \to T^*\cF\quad\mbox{,}\quad(q,p)\mapsto \big(q,p-p(N(q)) N^{\flat}(q)\big).\]
The cotangent bundle $T^*\cF$ carries the canonical symplectic form $\omega_\cF$.
\begin{TheoremN}  \label{thm:cool}\hspace*{-1mm}{\hypertarget{thmC}{\bf C}}
With respect to the embeddings $\imath_E^\pm:\cH_E^\pm \to T^*M$
one has 
\begin{equation}
\label{eq:symp:symp}
(\imath_E^\pm)^*\omega_0 = (n^\pm)^*\omega_F.
\end{equation}
So for the Riemannian volume element $d\cF$ induced by $h$ on $\cF\subseteq M$
\begin{equation}
\label{eq:Riemann:symp}
\int_{\cH_E^\pm}\frac{\omega_0^{\,\wedge n-1}}{(n-1)!} =v_{n-1}\int_\cF (2(E-V(q)))^{(n-1)/2}\,d\cF(q),
\end{equation}
with the Lebesgue measure $v_k$ of the $k$-dimensional unit ball.
\end{TheoremN}
\textbf{Proof:}
To prove \eqref{eq:symp:symp}, we represent $\cF$ as a zero set of a smooth function $f$, 
defined in a neighborhood $U\subseteq M$ of a point $q_0$ of $\cF$. 
We additionally require that $\bsd f(N(q))=+ 1$ $(q\in \widetilde{\cF}:=\cF\cap U)$. 
Next we multiply the phase space function $\pi^* f:T^*U\to \bR$ with the functions 
(depending on the parameter $E$) 
\[(q,p)\mapsto \mp \sqrt{2(E-V(q))-\|p\|_{h^{-1}}^2+p(N(q))^2}.\] 
The resulting functions $g^\pm:T^*U\to \bR$ 
define maximal Hamiltonian flows $\Gamma^\pm: D\to T^*U$. As $g^\pm$ vanishes on 
$T^*_{\widetilde{\cF}} M$, 
there the flow lines of $\Gamma^\pm$ coincide with those of the Hamilton flow of $f$. This means that
$\Gamma^\pm$ acts on the fibers of $\pi: T^*\widetilde{\cF}\to \widetilde{\cF}$:
\[\Gamma^\pm_t(q,p)= 
\big(q,p\,\mp\, \sqrt{2(E-V(q))-\|p\|_{h^{-1}}^2+p(N(q))^2} N^{\flat}(q) t\big) \qquad (q\in \widetilde{\cF}).\]
Note that $\Gamma^\pm_t$ does not change the argument of the square root.
For initial conditions $(q,p)\in \cH_E^\pm\cap T^*U$, this square root is the modulus of the (initial)
momentum component $p(N(q))$ in the normal direction.
So on $T^*U$ the flow $\Gamma^\pm_1$ maps $ \cH_E^\pm$ into the cotangent bundle $T^*\cF$.

Formula \eqref{eq:symp:symp} then follows like in Remark \ref{rem:additional}.3, using that 
$\omega_\cF= \imath_{T^*\cF}^* \, \omega_0$ for the embedding $ \imath_{T^*\cF}:T^*\cF\to T^*M$.

In the left hand side of \eqref{eq:Riemann:symp} we wrote for simplicity $\omega_0$ instead of
$(\imath_E^\pm)^*\omega_0$. When we apply \eqref{eq:symp:symp}, we obtain
\[\int_{\cH_E^\pm}\frac{((\imath_E^\pm)^*\omega_0)^{\wedge n-1}}{(n-1)!} = 
\int_{\cH_E^\pm}\frac{((n^\pm)^*\omega_F)^{\wedge n-1}}{(n-1)!} = 
\int_{n^\pm(\cH_E^\pm)}\frac{\omega_F^{\,\wedge n-1}}{(n-1)!} .\]
We write $\frac{\omega_F^{\,\wedge n-1}}{(n-1)!}$ in the form $d\cF\wedge d\cG$, where 
on the fiber  of $\pi: T^*\widetilde{\cF}\to \widetilde{\cF}$ over $q$, $d\cG_q$ denotes the Riemannian
volume element with respect to $h_q^{-1}$.
Restricted to the fiber over $q\in \widetilde{\cF}$ the image
$n^\pm(\cH_E^\pm)$ is a ball of radius $2(E-V(q))$ with respect to $\|\cdot\|_{h_q^{-1}}$.
Integrating out the fiber yields \eqref{eq:Riemann:symp}.
\hfill $\Box$\\[2mm]
Assume now that there is a $h$-orthogonal decomposition $M=M_1 \times M_2$ 
of configuration space, that is, for $Q=(Q_1,Q_2),\,Q'=(Q_1',Q_2')\in T_qM$
\[h_q(Q,Q')=h_{1,q}(Q_1,Q_1')+h_{2,q}(Q_2,Q_2').\]
This then extends to a $g$-orthogonal decomposition 
$T^* M = (T^*M_1) \times (T^*M_2)$ of the symplectic manifold $(T^*M,\omega)$. 
So  the Hamiltonian \eqref{Ham} has the form
\[H(q,p)=T_1(q,p_1)+T_2(q,p_2)+V(q_1,q_2)\qquad \big((q,p)=(q_1,p_1;q_2,q_2)\in T^*M\big),\]
with $T_i(q,p_i):= \eh h_{i,q}^{-1}(p_i,p_i)$.
For the foot point maps $\pi_i:T^*(M_i)\to M_i$, the volume form 
$\Omega:=\frac{\omega_i^{\wedge n}}{n!}$ on $T^* M$
can be written as $\Omega=\pi_1^*\Omega_1\wedge \pi_2^*\Omega_2$ w.r.t.\ the volume forms
$\Omega_i:=\frac{\omega_i^{\wedge n_i}}{n_i!}$ on $T^*M_i$.

Set $n_i:=\dim(M_i)$ so that $n=n_1+n_2$. We assume that a hypersurface $\cF\subseteq M$ 
has the property that both families
\[\cF^{\,q_2}_1:=\{q_1\in M_1\mid (q_1,q_2)\in \cF\}\qquad(q_2\in M_2)\]
and
\[\cF^{\,q_1}_2:=\{q_2\in M_2\mid (q_1,q_2)\in \cF\}\qquad(q_1\in M_1)\]
consist of hypersurfaces of $M_1$ respectively of $M_2$.\footnote{Locally this is a generic property.
If this is violated, then $\cF$ may still have this property outside a subset of measure zero.}  

\begin{Corollary}
Then, assuming finiteness of the integrals, $\int_{\cH^\pm_E} \frac{\omega^{\wedge n-1}}{(n-1)!}$ equals
\begin{align*}
&v_{n_2-1} \int_{T^*(M_1)} \left( \int_{\cF^{q_1}_2} 
(2(E-T_1(q,p_1)-V(q)))^{(n_2-1)/2} d\cF^{\,q_1}_2(q_2) \right) \Omega_1(q_1,p_1)\\
+\,
&v_{n_1-1} \int_{T^*(M_2)} \left( \int_{\cF^{q_2}_1} 
(2(E-T_2(q,p_2)-V(q)))^{(n_1-1)/2} d\cF^{\,q_2}_1(q_1) \right) \Omega_2(q_2,p_2).
\end{align*}
\end{Corollary}
\textbf{Proof:}
Using Theorem \hyperlink{thmC}{\bf C}, we have to prove that
\[\frac{\omega^{\wedge n-1}}{(n-1)!}=
\frac{\pi_1^*\omega_1^{\wedge n_1-1}}{(n_1-1)!}\wedge \Omega_2+
\frac{\pi_2^*\omega_2^{\wedge n_2-1}}{(n_2-1)!}\wedge \Omega_1.\]
But as $\omega=\pi_1^*\omega_1+ \pi_2^*\omega_2$ and $n-1-k=n_2+(n_1-1)-k$
\[\omega^{\wedge n-1} = \sum_{k=0}^n {n-1\choose k}\pi_1^*\omega_1^{\wedge k}\wedge \pi_2^*\omega_2^{\wedge n-1-k}
= \sum_{k=n_1-1}^{n_1} {n-1\choose k}\pi_1^*\omega_1^{\wedge k}\wedge \pi_2^*\omega_2^{\wedge n-1-k}.\]
Noting that 
\[\textstyle
 \frac{{n-1\choose n_1-1}}{(n-1)!}= \frac{1}{(n_1-1)!\,n_2!}\quad\mbox{and}\quad
 \frac{{n-1\choose n_1}}{(n-1)!}= \frac{1}{(n_2-1)!\,n_1!}\,,\]
we have proven the corollary.
\hfill $\Box$\\[2mm]

\section{Applications and Open Questions}
\label{sec:Applications}

In this paper, we have defined the set of transition points depending on a given sequence 
of Poincar\'e surfaces. 
In applications, the perspective is usually vice versa: one is interested in showing 
the improbability of a given subset of the wandering set. 
If the technique presented here is ought to be applied, the task then is to find an 
appropriate sequence of Poincar\'e surfaces, such that their total area decreases to zero 
and that the wandering orbits under considerations can be shown to be transition points to 
that sequence.

One important model, where these techniques can be implemented, is the $N$-body 
problem of celestial mechanics. A subset of singular orbits \eqref{def:sing}
which are of special interest are collisions: they are exactly those singular orbits, where 
every particle has a limit position in configuration space, as time approaches the singularity. 
In the upcoming paper \cite{Fleischer_Knauf_2018b}, we show the improbability of 
collisions in the $N$-body problem that may include some fixed centers as well. 
More generally, the result holds for a wide class of two-body interactions, including the
gravitational case of celestial mechanics, but also e.g.\ Coulomb fields in electrostatic motion.
Due to the technique devised in the present article, these estimates are optimal,
concerning the power law of the two-body interactions.

For the same family of two-body interactions, also non-collision singularities can be treated 
in certain situations: 
in the upcoming paper \cite{Fleischer_2018} we show the improbability of non-collision
singularities in systems with two free particles and an arbitrary amount of fixed centers. 
In the upcoming paper \cite{Fleischer_2018b} we show the improbability of non-collision 
singularities in the four-body system.

As pointed out, the technique presented here can only be used to show the improbability 
of certain subsets of wandering orbits. 
However, in principle the given result can be applied to show the improbability of certain 
sets of non-singular wandering orbits, e.g.\ those for which the asymptotic velocity does 
not exist as a limit.

\end{document}